\documentclass[aps,prl,reprint,superscriptaddress]{revtex4-1} 
\usepackage{amsfonts}
\usepackage{amsmath}
\usepackage{amssymb}
\usepackage{mathrsfs} 
\usepackage{mathrsfs}
\usepackage{color} 
\usepackage{graphicx}

\newcommand{\pd}[2]{\frac{\partial #1}{\partial #2}}

\begin{document} 
\title{Asymptotics of surface-plasmon redshift saturation at sub-nanometric separations}
\author{Ory Schnitzer}
\affiliation{\footnotesize{Department of Mathematics, Imperial College London, London SW7 2AZ, United Kingdom}}
\author{Vincenzo Giannini}
\affiliation{\footnotesize{The Blackett Laboratory, Department of Physics, Imperial College London,  London SW7 2AZ, United Kingdom}}
\author{Richard V.~Craster}
\affiliation{\footnotesize{Department of Mathematics, Imperial College London, London SW7 2AZ, United Kingdom}}
\author{Stefan A.~Maier}
\affiliation{\footnotesize{The Blackett Laboratory, Department of Physics, Imperial College London,  London SW7 2AZ, United Kingdom}}
\begin{abstract}
Many promising nanophotonics endeavours hinge upon the unique plasmonic properties of nanometallic structures with narrow non-metallic gaps, which support super-concentrated bonding modes that singularly redshift with decreasing separations. In this letter, we present a descriptive {physical} picture, complemented by elementary asymptotic formulae, of a nonlocal mechanism for plasmon-redshift saturation at subnanometric gap widths. Thus, by considering the electron-charge and field distributions in the close vicinity of the metal-vacuum interface, we show that nonlocality is asymptotically manifested as an effective potential discontinuity. For bonding modes in the near-contact limit, the latter discontinuity is shown to be effectively equivalent to a widening of the gap. As a consequence,    
 the resonance-frequency near-contact asymptotics are a renormalisation of the corresponding local ones. Specifically, the renormalisation furnishes an asymptotic plasmon-frequency lower bound that scales with the $1/4$-power of the Fermi wavelength. 
We demonstrate these remarkable features in the prototypical cases of nanowire and nanosphere dimers, showing agreement between our elementary expressions and previously reported numerical computations. 
\end{abstract}

\maketitle

\textit{Introduction}.---There is currently great interest in the underlying physics and nanotechnological applications of plasmonic light confinement in gaps separating nearly touching metal nanostructures down to nanometric separations and below \cite{Scholl:12, Ciraci:12, Savage:12}. 
For such close separations, the ``bonding'' surface-plasmon modes redshift with decreasing separations and become highly localised \cite{Hill:10,Aubry:10A,Lei:12}. Theoretical approaches going beyond the classical electromagnetic formulation have included \textit{ab initio} quantum-mechanical formulations \cite{Zuloaga:09, Marinica:12}, and phenomenological hydrodynamic \cite{Mcmahon:09,Ciraci:13,Mortensen:14,Toscano:15,Raza:15} and quantum-corrected models \cite{Esteban:12}. In particular, numerical simulations and approximate analytic analyses of the hydrodynamic Drude model \cite{Garcia:08, Fernandez:12, Ciraci:13b,Luo:13,Raza:15b}, which in many cases qualitatively captures nonlocality \cite{Stella:13},
reveal that the plasmon-frequency redshift saturates with decreasing particle separation; this is not captured by a purely local description of the dielectric response, yet can be relatively easily inferred via monitoring light scattering from a metal nano particle in near-contact with a plane substrate \cite{Ciraci:12} --- a surprising far-field optical manifestation of a quantum effect. We here apply scaling and asymptotic arguments to arrive at an intuitive physical picture, complemented by asymptotic formulae, of a nonlocal mechanism underpinning the redshift saturation.  


\textit{Local description}.---We begin with a discussion of the surface-plasmon redshift in the local approximation. For specificity, consider a dimer nano-metallic structure comprised of two identical convex metal particles surrounded by vacuum {(characteristic particle size $a$, minimal separation $d$)}. In the quasi-static approximation \cite{Maier:07}, the electric near-field is derived from potentials $\bar\varphi$ and $\varphi$ respectively satisfying Laplace's equation within and outside the metal, along with the interfacial conditions 
\begin{equation}\label{local bc}
\varphi=\bar\varphi, \quad \pd{\varphi}{n}=\epsilon\pd{\bar\varphi}{n},
\end{equation}
where $\partial/\partial n$ denotes the outward normal derivative, and $\epsilon$ is the relative dielectric function of the metal. A surface-plasmon mode with eigenvalue $\epsilon$ corresponds to a non-trivial solution of the above problem with $\varphi$ attenuating at large distances. The plasmon frequencies $\omega$ are obtained by inverting $\epsilon(\omega)$; assuming a Drude metal,
\begin{equation}\label{Drude}
\epsilon(\omega) = 1-\frac{\omega_p^2}{\omega^2+i\gamma\omega},
\end{equation}
where $\omega_p=e\sqrt{\mathcal{N}_e/\epsilon_0 m}$  is the plasma frequency ($e$ being the electron-charge magnitude, $\mathcal{N}_e$ the equilibrium electron density, and $m$ the effective electron mass) and $\gamma$ the collision frequency. Strictly speaking, since $\epsilon$-eigenvalues are real and negative \cite{Mayergoyz:05}, surface modes exist when $\gamma=0$ (and $\omega<\omega_p$). For $0<\gamma\ll\omega$, however, a damped resonance occurs for external forcing having appropriate symmetries, wherein surface modes are excited with an amplification factor inversely proportional to $\mathrm{Im}[\epsilon]\approx \omega_p^2\gamma/\omega^3$. 

\begin{figure}[t]
\begin{center}
{\includegraphics[width=8.5cm]{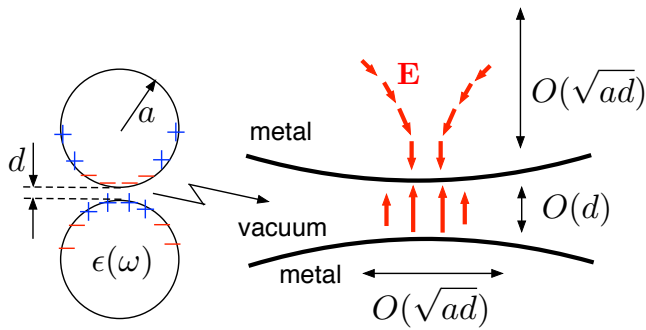}}
\caption{Schematics of the fundamental axisymmetric ``bonding'' mode of a metal nanosphere dimer in the local approximation. The inset zooms in on the gap and metal-pole regions wherein the potential varies rapidly. The plasmon eigenvalue $\epsilon$ is negative and $O(\sqrt{d/a})$-large in order to allow displacement continuity.}
\label{gap}
\end{center}
\end{figure}

The symmetry of the dimer configuration implies that surface-plasmon eigen-potentials are either symmetric or anti-symmetric about the plane bisecting the gap.  Our interest is in the anti-symmetric ``bonding'' modes, where the polarisation-charge distribution on one side of the gap is opposite to that on the other, implying a gap field that becomes dominantly transverse with decreasing separation (see Fig.~\ref{gap}). Henceforth we shall be interested in the asymptotic near-contact limit where $d/a\ll1$. Given the local paraboloidal (or parabolic, in two dimensions) geometry of the gap, the potential $\varphi$ rapidly varies in the transverse direction over distances $O(d)$ --- approximately linearly, as a leading-order balance of Laplace's equations reveals --- and in directions tangent to the bisecting plane over distances $O(\sqrt{a d})$. Continuity in potential across the gap boundary suggests that $\bar\varphi$ too varies rapidly over $O(\sqrt{ad})$ tangential distances. Owing to the symmetry of the Laplace operator, and the apparent unboundedness of the metal domain away from the gap boundary on such small scales, $\bar\varphi$ must vary over comparable distances also in the transverse direction. 

The above description implies that bonding modes have their field more strongly localised on the vacuum side than on the metal side; it is this property that allows a strong redshift as $d/a\to0$. In fact, we readily deduce a scaling relation in this limit by invoking electric-displacement continuity. If the potential in the gap region is $O(\varphi_g)$, say, then the transverse field there is $O(\varphi_g/d)$ whereas in the metal the field is $O(\varphi_g/\sqrt{ad})$. The scaling result (see also Refs.~\onlinecite{Lebedev:10} {and \onlinecite{Schnitzer:15arxiv}})
\begin{equation}\label{eps scaling}
\epsilon(\omega_{\text{res}}) \sim - {\alpha}(d/a)\left(\frac{a}{d}\right)^{1/2} \quad \text{as} \quad d/a\to0,
\end{equation}
then follows, where $\alpha>0$ is an $O(1)$ pre-factor dependent on the specific geometry and mode number; henceforth we shall simply write $\epsilon$ for $\epsilon(\omega_{\text{res}})$. More subtle scaling considerations show for planar cases that $\alpha$ is a constant governed solely by the local gap morphology, whereas in three dimensions $\alpha$ is a weak logarithmic function of $d/a$ that additionally depends on certain integral features of the particle-scale geometry
\footnote{{Consider the zero-net-charge constraint $\oint \partial{\varphi}/{\partial n}\,dA=0$,  the integral taken over a particle surface. The contribution to this integral from the vicinity of the gap is estimated as the gap field $O(\varphi_g/d)$ times the area of integration: $O(\sqrt{ad})$ per unit length in two dimensions and $O(ad)$ in three dimensions. Similarly, the particle-scale contribution is $O(\varphi_g/a)$ times the area: $O(a)$ per unit length in two dimensions and $O(a^2)$ in three dimensions. Consequently, the gap and particle-scale domains are to leading order decoupled (coupled) in the planar (three dimensional) case. In three dimensions the gap charge must counterbalance the particle-scale charge associated with the capacitance of two nearly touching approximately equipotential (since $|\epsilon|\gg1$) particles, which is well known from electrostatics to be logarithmically singular in $d/a$ \cite{Jeffrey:78}.}}. 

In the case of a circular nano-wire dimer (radius $a$), 
\begin{equation}\label{alpha wires}
\alpha = \frac{1}{n+1}, \quad n=0,1,2,\ldots
\end{equation}
One way to derive \eqref{alpha wires} is by asymptotically reducing exact analytical solutions based on transformation optics \cite{Aubry:10A} or separation of variables in bi-polar coordinates \cite{Klimov:14} (see e.g.~Ref.~\onlinecite{Vorobev:10}). A powerful alternative developed in Ref.~\onlinecite{Schnitzer:15arxiv}, not requiring an analytic solution, is an \textit{ab initio} singular perturbation analysis of the plasmonic eigenvalue problem in the near-contact limit; the surface-plasmon modes emerge by matching relatively simple leading terms of spatially overlapping asymptotic expansions and the method is essentially a detailed version of the scaling arguments given earlier. 
{In the significantly more complicated case of a sphere dimer this approach yields \cite{Schnitzer:15arxiv}}
\begin{equation} \label{alpha spheres}
\alpha = \frac{{2}}{2\tilde{n}(n,\ln (d/a))+1}, \quad n=0,1,2,\ldots
\end{equation}
{(axisymmetric bonding modes); for each $n=0,1,2...$,  $\tilde{n}(n,\ln (d/a))$ is the root of the transcendental equation $2\Psi(-\tilde{n})=\ln(a/4d)$ that approaches $n$ as $d\to0$ \footnote{The Digamma function is singular at non-positive integer values.},
 with $\Psi$ being the Digamma function \cite{Abramowitz:book}.}

\textit{Nonlocal description}.---For a Drude metal \eqref{Drude}, the singular scaling \eqref{eps scaling} implies a plasmon-frequency redshift such that $\omega/\omega_p=O\left((d/a)^{1/4}\right)$. The local model \eqref{local bc} underlying \eqref{eps scaling} breaks down, however, when $d$ becomes comparable to the sub-nanometric Fermi wavelength $\lambda=\beta/\omega_p$ ($\beta$ is a parameter characterising nonlocality; for $\omega\gg\gamma$, $\beta$ is $\sqrt{3/5}$ times the Fermi velocity \cite{Raza:15}). The finite dispersion of the electron-density deviation from the equilibrium value $\mathcal{N}_e$ becomes important, and we explore how the asymptotic result \eqref{eps scaling} is modified by invoking the hydrodynamic Drude model; the electron-density deviation $\mathcal{N}$ is solved for in conjunction with the potential $\bar\varphi$ within the metal, which is now associated with a microscopic rather than a macroscopic electromagnetic field. As reviewed in Ref.~\onlinecite{Raza:15}, the governing equations in the metal are obtained by linearising the electron-density continuity equation and a Navier--Stokes equation governing the electron-gas flux per unit density. In the quasi-static approximation, and suppressing a time harmonic dependence of the form $\exp(-i\omega t)$, one finds:
\begin{equation}\label{nonlocal model}
\lambda^2 \nabla^2 \mathcal{N} = \frac{\epsilon}{\epsilon-1} \mathcal{N}, \quad \epsilon_0\nabla^2\bar\varphi = e\mathcal{N},
\end{equation}
where $\epsilon$ is given by \eqref{Drude}. 
The vacuum potential $\varphi$ is still governed by Laplace's equation. The model is supplemented by the interfacial conditions
\begin{equation}
\varphi=\bar\varphi, \quad \pd{\varphi}{n}=\pd{\bar\varphi}{n}, \quad \frac{m\beta^2}{e\mathcal{N}_e}\pd{\mathcal{N}}{n}=\pd{\varphi}{n},
\end{equation}
the first two being the standard microscopic electromagnetic conditions, while the third {dictates} a vanishing electron-density flux through the interface ({neither electron spill-out nor tunnelling are accounted for}).

The nonlocal model is apparently a significant complication, but the disparity between the sub-nanometric Fermi-wavelength $\lambda$ and the characteristic scale $a$ can be exploited. For $\omega<\omega_p$, $\epsilon$ is negative and the first of \eqref{nonlocal model} is a modified Helmholtz equation, implying that electron-density deviations decay exponentially over an $O(\lambda)$ distance away from the interface. In the bulk metal outside the latter electron-charge ``boundary layers'', $\mathcal{N}/\mathcal{N}_e$ is exponentially small in $\lambda/a$, whereby the second of \eqref{nonlocal model} reduces to Laplace's equation. This suggests carrying out a zoomed-in analysis of the boundary layers to arrive at an effective coarse-grained model relating the vacuum and bulk-metal potentials. 

\begin{figure}[t]
\begin{center}
{\includegraphics[width=6.5cm]{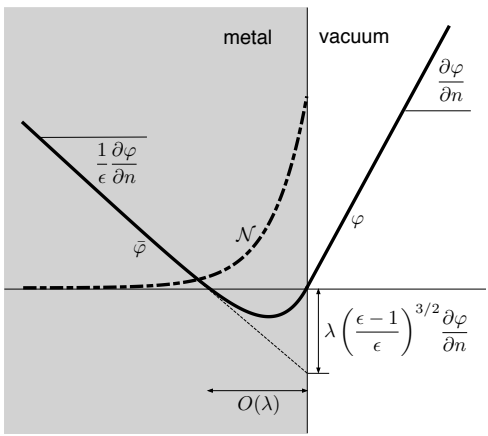}}
\caption{Quasi-one-dimensional potential and electron-density perturbation profiles in the vicinity of a metal-vacuum interface. The electron-charge boundary layer can be effectively replaced by a potential jump proportional to the interfacial normal field.}
\label{intuit}
\end{center}
\end{figure}

Before we return to discuss the near-contact limit, let us consider for a moment a smooth metal particle or structure characterised by a \emph{single} length scale. It is then evident that changes in potential over the interface are slow relative to the rapid variations in the boundary layer. Further assuming a locally flat interface, an approximate solution of \eqref{nonlocal model} is readily obtained as
\begin{multline}\label{oned}
\frac{\mathcal{N}}{\mathcal{N}_e}\approx \frac{e\lambda }{m\beta^2}\pd{\varphi}{n}\left(\frac{\epsilon-1}{\epsilon}\right)^{1/2}\exp\left[{\left(\dfrac{\epsilon}{\epsilon-1}\right)^{1/2}\dfrac{x}{\lambda}}\right], \\
\bar\varphi-\varphi \approx  \frac{x}{\epsilon}\pd{\varphi}{n}  \\ + \lambda\left\{\exp\left[{\left(\dfrac{\epsilon}{\epsilon-1}\right)^{1/2}\dfrac{x}{\lambda}}\right]- 1\right\}\left(\frac{\epsilon-1}{\epsilon}\right)^{3/2}\pd{\varphi}{n},
\end{multline}
where $x$ denotes distance from the interface ($x>0$ is vacuum), and $\varphi$ and $\partial{\varphi}/\partial{n}$ are evaluated at the interface (see Fig.~\ref{intuit}). 
The quasi-one-dimensional solution \eqref{oned} implies that non-locality can be approximately accounted for through effective interfacial conditions,
\begin{gather}\label{nonlocal full}
\varphi-\bar\varphi \sim \lambda\left(\frac{\epsilon-1}{\epsilon}\right)^{3/2}\pd{\varphi}{n}, \quad \epsilon\pd{\bar\varphi}{n}\sim\pd{\varphi}{n},
\end{gather}
where $\bar\varphi$, now representing the bulk-metal potential, satisfies Laplace's equation. For the  single-scale particle, the effective potential discontinuity in \eqref{nonlocal full} results in a small $O(\lambda/a)$ correction to $\epsilon$, corresponding to a small $O(\lambda/a)$ blueshift of $\omega/\omega_p$ from values based on the local model. A more systematic derivation of \eqref{nonlocal full} based on the method of matched asymptotic expansions {\cite{Hinch:91}}, for a generic smoothly curved interface, reveals that \eqref{nonlocal full} is indeed asymptotic to $O(\lambda/a)$ but is corrected by additional effective terms at $O(\lambda^2/a^2)$.

\begin{figure}[b]
\begin{center}
{\includegraphics[width=5.5cm]{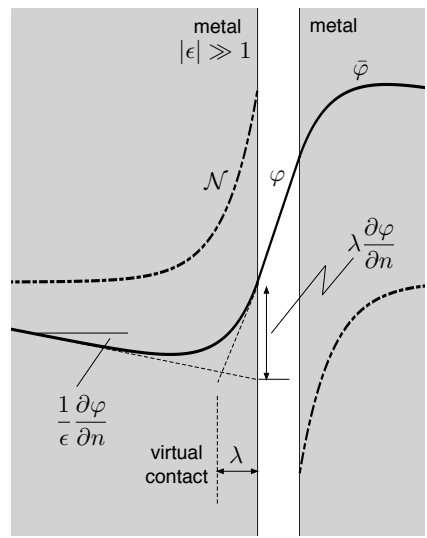}}
\caption{{Quasi-one-dimensional electron-density perturbation and potential profiles characteristic of a gap bonding mode. When the separation is $O(\lambda)$, nonlocality is appreciable; the macroscale boundary conditions \eqref{nonlocal gap} governing the leading-order large eigenvalue $-\epsilon$ feature an effective potential jump that can {alternatively} be interpreted as {a transverse inward} shift {by $\lambda$} of the metal boundary.}}
\label{intu_gap}
\end{center}
\end{figure}

We now return to the case of a dimer in the near contact limit $d/a\ll1$. Evidently, for $d\gg \lambda$, the local picture is approximately correct. When $d=O(\lambda)$, however, nonlocality is an appreciable effect. Indeed, we previously saw that the transverse field in the gap is $O(\varphi/d)$-large, whereby the  potential jump suggested by \eqref{nonlocal full} is upgraded to leading-order status. In studying the near-contact limit we shall focus on the leading-order behaviour, disregarding small corrections; then, the quasi-one-dimensional solution \eqref{nonlocal full} remains relevant towards describing the gap region (while tangential variations are rapid in the near-contact limit, they are still less rapid than the transverse variations in the charge boundary layer). Subject to a \textit{posteriori} verification, we still   expect $|\epsilon|\gg1$, hence a uniformly valid leading-order macroscale model reads 
\begin{equation}\label{nonlocal gap}
\varphi-\bar\varphi \sim \lambda\pd{\varphi}{n}, \quad \epsilon\pd{\bar\varphi}{n}\sim\pd{\varphi}{n}.
\end{equation}
Note that, away from the gap, the potential varies over $O(a)$ distances and hence the effective potential jump is negligible in the present leading-order scheme. 

We can now make a key observation: The effective model \eqref{nonlocal gap} is to leading order equivalent to a shift of the metal boundary by $\lambda$ in the inward normal direction. {In fact, since in the vicinity of the gap the normal vector is approximately transverse, the nonlocal effect can be interpreted simply as a widening of the gap, see Fig.~\ref{intu_gap}.} 
Remarkably, {these asymptotic analogies} imply that the nonlocal $\epsilon$-asymptotics are simply a renormalisation of the local result \eqref{eps scaling}:
\begin{equation}\label{renorm}
\epsilon \sim - \alpha\left(\frac{d+2\lambda}{a}\right) \left(\frac{a}{d+2\lambda}\right)^{1/2}, 
\end{equation}
where $\alpha$ is the same $O(1)$ constant (in 2d) or logarithmic function (in 3d) as in the local theory. 
We note that \eqref{renorm} can be alternatively derived, {in the spirit of Ref.~\onlinecite{Schnitzer:15arxiv},} by a direct singular perturbation analysis of the original nonlocal model \eqref{nonlocal model} or either of the reduced models \eqref{nonlocal full},\eqref{nonlocal gap}.
\begin{figure}[t]
\begin{center}
{\includegraphics[width=9.5cm]{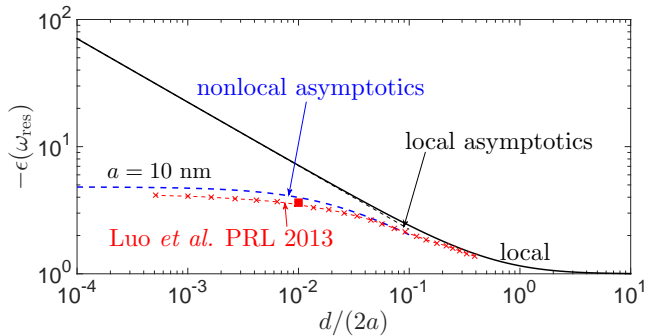}}
\caption{
Resonant-frequency condition for the fundamental bonding mode of a gold circular nanowire dimer of radius $a$, Drude dielectric function $\epsilon(\omega)$ [cf.~\eqref{Drude}], and Fermi wavelength $\lambda=0.215$ nm. Black solid: local analytical solution \cite{Klimov:14}. Black dashed:  local near-contact asymptotics \eqref{eps scaling} \& \eqref{alpha wires}. Blue dashed: renormalised nonlocal near-contact asymptotics \eqref{renorm}. Red square: fully retarded nonlocal simulation reported in Ref.~\onlinecite{Luo:13}. Red dashed: calculations for fitted dielectric-layer model reported in Ref.~\onlinecite{Luo:13}.}
\label{fig:wires}
\end{center}
\end{figure}

\begin{figure}[t]
\begin{center}
{\includegraphics[width=9.5cm]{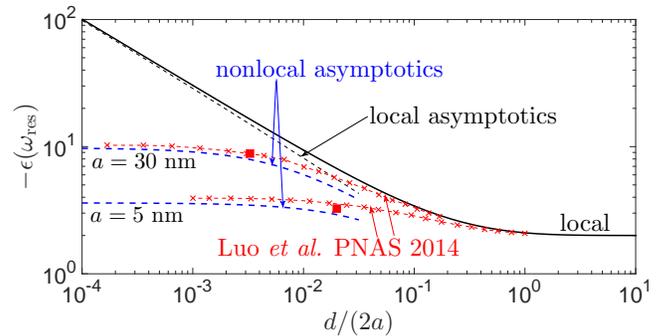}}
\caption{Same as Fig.~\ref{fig:wires} but for a gold nanosphere dimer. Black solid: local semi-numerical solution \cite{Schnitzer:15arxiv}. Black dashed:  local near-contact asymptotics \eqref{eps scaling} \& \eqref{alpha spheres}. Blue dashed: renormalised nonlocal near-contact asymptotics \eqref{renorm}. Red square: fully retarded nonlocal simulation reported in Ref.~\onlinecite{Luo:14}. Red dashed: calculations for modified dielectric-layer model reported in Ref.~\onlinecite{Luo:14}.}
\label{fig:spheres}
\end{center}
\end{figure}

Predictions of Eq.~\eqref{renorm} for the zeroth-mode of gold nanowire and nanosphere dimers are depicted by the blue dashed lines in Fig.~\ref{fig:wires} and \ref{fig:spheres}, respectively. Also shown: (i) Black solid lines ---  local theory (analytic solution for wires \cite{Klimov:14} and a semi-numerical method for spheres \cite{Schnitzer:15arxiv}); (ii) Black dashed lines --- local near-contact asymptotics \eqref{eps scaling}; (iii) Numerical results reported in Refs.~\onlinecite{Luo:13,Luo:14} --- Red square symbols represent a fully retarded nonlocal simulation whereas the red dashed lines are fitted solutions of a reduced dielectric-layer model, which we shall discuss later.  There is satisfying agreement with the numerical data of Refs.~\onlinecite{Luo:13,Luo:14}, which smoothly connect the local approximations at $d\gg\lambda$ with our renormalised asymptotics at $d=O(\lambda)$; especially given the $O(1)$ error, small compared to $\sqrt{a/d}$, involved in \eqref{renorm}.

The saturation of $\epsilon$ observed in Fig.~\ref{fig:wires} and \ref{fig:spheres} is evident in \eqref{renorm}. Indeed, for $d\ll\lambda$ we find
\begin{equation} \label{sat eps}
\epsilon \sim - \alpha(2\lambda/a)  \left(\frac{a}{2\lambda}\right)^{1/2}.
\end{equation}
Formally, $\omega/\omega_p\gg1$ when $|\epsilon|\gg1$, and hence the leading-order asymptotic saturation frequency corresponding to \eqref{sat eps} is
\begin{equation}\label{omega}
\omega/\omega_p\sim (2\lambda/a)^{1/4}/[\alpha(2\lambda/a)]^{1/2}.
\end{equation}
For example, in the case of a circular nanowire dimer, substituting \eqref{alpha wires} into \eqref{omega} furnishes the asymptotic lower bound $\omega/\omega_p\sim (2\lambda/a)^{1/4}$. 

\textit{Luo {et al.}'s dielectric-layer model}.---We have shown that nonlocality can be asymptotically accounted for in terms of an effective potential jump, and in some cases also as an inward shift of the metal-vacuum boundary. {Previous local-analogue models \cite{Fernandez:12,Luo:13,Luo:14}, underlined by physically intuitive rather than asymptotic arguments, involve a degree of freedom thereby allowing to fit numerical computations}. In particular, Luo \textit{et al.} \cite{Luo:13} suggested to account for nonlocality by replacing a thin layer of metal {of thickness $l_d$} adjacent to the interface with a dielectric material of relative permittivity $\epsilon_d$. By requiring the reflection and transmission coefficients in a planar geometry to be consistent with a complete nonlocal solution, Luo \textit{et al.} derived an expression for the ratio $l_d/\epsilon_d$, which in our notation reads 
\begin{equation} \label{dielectric}
l_d/\epsilon_d = \lambda\left(\frac{\epsilon-1}{\epsilon}\right)^{3/2}.
\end{equation}

Under certain restrictions, condition \eqref{dielectric} is consistent with our model \eqref{nonlocal full}, and with \eqref{nonlocal gap} in particular. 
To see this, consider yet again our quasi-one-dimensional approximation, this time for a triple-layer system comprised of a local metal ($x<-l_d$), a dielectric layer ($-l_d<x<0$) and vacuum ($x>0$). It is immediate to write down equations connecting the potentials and fields at the boundaries of the dielectric layer:
\begin{equation}
\epsilon\left.\pd{\bar\varphi}{x}\right|_{-l_d}\approx \left.\pd{\varphi}{x}\right|_{0}, \quad \left.{\varphi}\right|_{0} - \left.{\bar\varphi}\right|_{-l_d}\approx \frac{l_d}{\epsilon_d}\left.\pd{\varphi}{x}\right|_{0}.
\end{equation} 
It may naively appear that by substituting \eqref{dielectric} we find the asymptotic macroscale model \eqref{nonlocal full} for any fitting choice. It should be noted, however, that while in the asymptotic model  \eqref{nonlocal full} the conditions apply at the true interface, here the metal variables are evaluated at a distance of $l_d$ from the interface. For single-length-scale structures this generally carries an $O(\lambda/a)$ relative error, whereas the error in \eqref{nonlocal full} is smaller. This implies the fitting constraint $l_d\ll \lambda$, which might explain why in Ref.~\onlinecite{Luo:13} fixing $\epsilon_d=1$ did not give a good prediction of high-frequency surface-plasmon polaritons  (as $\omega\to\omega_p$,  $\epsilon$ becomes small and hence $l_d\gg\lambda$). This constraint is less important in the near-contact scenario, where the effect of nonlocality is dominant and {the error associated with the small geometric shift is negligible}. Indeed, fixing $\epsilon_d=1$ gave rather good results in Ref.~\onlinecite{Luo:13} {for nearly touching nanoparticles}, especially around the {fundamental} plasmon frequency. These remarks are consistent with the discussion in Ref.~\onlinecite{Luo:14}, where a modification of \eqref{dielectric} involving a non-uniform {dielectric-layer} thickness is employed. 

\textit{Conclusions}.---This letter provides an intuitive description of a nonlocal mechanism of redshift saturation captured by an elementary renormalisation of the local near-contact asymptotics. For a general metal nanostructure, with small separations, all of the theory presented herein can be adopted with minor changes related to having different metals and the identification of the bonding modes. 

The authors acknowledge funding from the Engineering and Physical Sciences Research Council (EPSRC) programme grant EP/L024926/1 Mathematical Fundamentals of Metamaterials.

\bibliography{refs.bib}
\end{document}